\documentstyle[preprint,aps]{revtex}

\input epsf.sty

\begin{document} \tolerance 50000

\draft

\title{The Short Range RVB State of Even Spin Ladders: \\
A Recurrent Variational Approach} 
\author{Germ{\'a}n Sierra$^{1}$ and
 Miguel A. Mart{\'\i}n-Delgado$^{2}$ } 
\address{ $^{1}$Instituto de Matem{\'a}ticas y
F{\'\i}sica Fundamental, C.S.I.C.,Madrid, Spain.
\\ 
$^{2}$Departamento de
F{\'\i}sica Te{\'o}rica I, Universidad Complutense. Madrid, Spain
 \\
}

\maketitle \widetext

\vspace*{-1.0truecm}

\begin{abstract} \begin{center} \parbox{14cm}{
Using a recursive method we construct dimer and nondimer 
variational ansatzs of the ground state for the two-legged ladder,
and compute the number of dimer coverings, the
energy density and the spin correlation functions.
The number of dimer coverings are given by the Fibonacci numbers
for the dimer-RVB state  and their generalization for the nondimer ones.
Our method relies on the  recurrent
relations satisfied by  the overlaps of the
states with different lengths, which can be solved 
using generating functions. 
The  recurrent relation method is applicable to other
short range systems. Based on our results we make a conjecture
about the bond amplitudes of the 2-ladder.}
\end{center}
\end{abstract}

\pacs{ \hspace{2.5cm} PACS numbers: 05.50.+q, 75.10.-b, 75.10.Jm}

\narrowtext

\section{Introduction}

There is an increasing theoretical and experimental 
interest in systems formed by
a finite number $n_l$ of coupled chains, currently known as 
ladders (for a review see \cite{DR}).
Theoretical studies suggested that antiferromagnetic spin ladders 
should be gapped
(gapless) depending on whether $n_l$ is an even (odd) number 
\cite{DRS},\cite{SM},\cite{Barnes}, \cite{RGS}, \cite{PSZ},
\cite{wns},\cite{GRS}.
This prediction has been confirmed experimentally in compounds 
like (VO)$_2$P$_2$O$_7$
($n_l = 2$) \cite{VOPO}, although some doubts has been recently casted
on the ladder structure on this material \cite{Garret}
, SrCu$_2$O$_3$ ($n_l=2$) \cite{SrCu}, 
Sr$_2$Cu$_3$O$_5$ ($n_l=3$), etc.
Moreover, the theory also predicts that upon doping the even-$n_l$ ladders
should become superconductors due to 
the pairing of holes \cite{DRS},\cite{RGS},\cite{NWS}. 
The recent discovery of superconductivity in the 2-legged ladder compound 
Sr$_{0.4}$Ca$_{13.6}$Cu$_{24}$O$_{41}$ under high pressure \cite{U}
may perhaps constitute a confirmation of this prediction \cite{BS}.
On the other hand the odd-$n_l$ ladders are not expected to superconduct.

Altogether the ladders fall in two different 
universality classes depending on the
even/oddness of $n_l$, which in the 
limit $n_l \rightarrow \infty$ should converge to the
same class. An example of this is given by the behaviour of the spin gap of the
even-$n_l$ ladders which vanishes exponentially with $n_l$ 
\cite{GS}, \cite{Ch},\cite{GCh}.

A field theoretical characterization of 
these two universal classes can be obtained
by mapping the spin ladders into the $1+1$ 
non-linear sigma model \cite{Sen}, \cite{GS}. The value
of the coupling constant $\theta$, 
which multiplies the instanton number in the action, is given by 
$\theta = 2 \pi S n_l$, 
where $S$ is the spin
of the chain \cite{Kh}, \cite{GS}, \cite{Mo}. For $S =1/2$ and 
$n_l$ even one gets $\theta=0$ (mod 2$\pi$),
which corresponds to a sigma model with a dynamically  generated gap,
while for $n_l$ odd one gets $\theta = \pi$ (mod 2$\pi$), 
which corresponds to a
gapless sigma  model which flows under the RG
to the $SU(2)$ level 1 Wess Zumino
model\cite{HA}. 
Thus the behaviour of ladders parallels that of the spin chains
as function of the spin, as first cojectured by Haldane using precisely
the mapping of the spin chains into the non-linear-sigma 
model \cite{H}, \cite{Aff}.

There is an alternative explanation of the qualitative 
difference between the even/odd
ladders based on the  
RVB (Resonating Valence Bond) theory.
\noindent
According to the authors of ref. \cite{wns}, the even-$n_l$ 
ladders are short ranged
RVB systems with confinement of spin deffects, which leads to the
existence of a 
spin gap  and exponential decaying correlation functions; 
while the odd-$n_l$ ladders are long-range RVB systems with no
confinement of deffects, no gap and power-law correlations. 
Strictely speaking, the 
above RVB interpretation based on the original ideas of 
Liang et al. \cite{LDA} concerning the 2D AF-Heisenberg system, are an
intuitive 
picture to explain the numerical 
results obtained using the DMRG \cite{DMRG} (Density Matrix
Renormalization Group). 
It is therefore interesting  to test the RVB picture using different
techniques in order to confirm and  get further insights into the short range
nature of the even legged-ladders.

The first step in this direction 
is the study of the dimer-RVB state (also termed NNRVB, 
standing for nearest-neighbour RVB). This has already been pursued in 
\cite{fan-ma},\cite{shapir},\cite{wns} for $n_l = 2$. Using the DMRG, 
the main conclusion of \cite{wns} 
is that one has to consider valence bonds between 
sites which are not NN (nearest-neighbours), 
going in that way beyond the dimer-RVB
towards a short-range RVB state, 
whose structure has not yet been studied in detail.

From a mathematical viewpoint the 
dimer-RVB states are relatively easy to handle
due to the work of Sutherland \cite{sutherland} 
which gives an elegant  diagrammatic
way of computing overlaps between 
the dimer states which form
the dimer-RVB state. 
However the  combinatorics of the
nondimer-RVB states has not yet been worked out and
their construction seems a priori a difficult task.
We shall show in this paper that there is a way 
to circumvent the computational problem associated 
with short-range
RVB states, which is based on 
the use of recurrent relations.
The main idea is to build up the ground state of a ladder
with $N + \nu$ rungs using the knowledge of  the ground states
with $N, N+1, \dots , N+\nu-1$ rungs.
This is achieved by a recurrent relation (RR) which gives
the ground state  $|N + \nu \rangle$ in terms of the ground
states
$ |N + \nu -1 \rangle,
\dots,  |N \rangle $. We shall call $\nu$ the order of the recurrent
relation. The ``matching" of the various g.s.'s within the
RR  is achieved by means of
a collection of $\nu$ states $ |\phi_1 \rangle,
 \dots |\phi_\nu \rangle$,  which are the elementary building 
blocks of our method. Loosely speaking,  $|\phi_k \rangle $
is a state which contains at least a  bond of length  $k$. 
For even-$n_l$ ladders, with $\xi$ a few lattice spacings,
we expect an adequate description of 
the ground state for small values of $\nu$. 
The use of RR's  allow also the
introduction, in a natural fashion, 
of variational parameters which we can determine
by the standard minimization procedure of the ground state energy.
The whole approach based on RR's 
is analytic, powerful and rather straightforward,
and we believe it 
represents a significant methodological improvement
as compared with previous analytic 
approaches to the study of RVB states, which
is applicable to other type of short range  systems.

The organization of the paper is as follows.
In section II we introduce the recurrent relations and compare the
states generated by them with the conventional RVB states.
In section III we deduce the RR's  for  the 
norm  of the ground state and the expectation value of the Hamiltonian.
These RR's are solved in section IV using  generating 
functions, which
allow us to obtain quite easily an analytic formula for the ground  
state energy density. The results of the
minimization of the ground state energy respect to the 
variational parameters of the ansatzs corresponding to $\nu$ = 2 and 3,
are presented in section V. In section VI we derive analytic
expressions for the correlation length $\xi$ and give their numerical
values for $\nu =$ 2 and 3. Finally in section VII
we discuss our results and the perspectives of the RR method.

\newpage

\section{RVB and RVA States}

Generically, an RVB state is a linear 
superposition of singlets states constructed
by pairing up the $N$ spins of a system into ${N\over 2}$ bonds ($N$ is even).
Making an analogy with the BCS states one can think of an amplitude $h(i-j)$
of creating a bond, denoted by $(ij)$,  
between the sites $i$ and $j$ \cite{An}. We shall work  with bipartite 
lattices and such that $i$ $(j)$ belongs to the sublattice
A (B).  The total RVB wave funtion is given by \cite{LDA},

\begin{equation}
|\psi \rangle_{RVB} = \sum_{i\in A, j\in B} h(i_1 - j_1) \cdots 
h(i_{N/2} - j_{N/2}) \ (i_1 j_1) \cdots (i_{N/2} j_{N/2}) 
\label{1}
\end{equation}

\noindent
where the amplitudes $h(i-j)$ are all positive or zero in order
to satisfy the Marshall sign rule \cite{Mar}. 
For a dimer state $h(i-j) \neq 0$
unless $i$ and $j$ are nearest neighbours (NN). 
In a short-range RVB state, $h(i-j)$
decreases exponentially with the distance while  
in a long-range RVB state, $h(i-j)$
decays algebraically as $|i-j|^{-p}$ for some power $p$.

Let us consider the RVB state (\ref{1}) 
for the two-leg ladder depicted in figure 1.
We have labeled the sublattice A by $k=1,2, \ldots , N$ and the sublattice B
by $\bar{k} =\overline{1}, \overline{2}, \ldots, \overline{N}$. 
The general ansatz (\ref{1}) only depends 
on the amplitudes $h_l$ of the bond 
$(k,\overline{k+l})$ for $l= 0, 1, \dots$. 
Let us consider some examples
in order to get familiar with the notation, 
while going through the details of the formalism.

\subsection{Columnar State}

\indent
Let us first assume  that only $h_0$ is non zero.
We normalize it  as
$h_0=1$. In this case the corresponding RVB state (\ref{1})
for a ladder with $N$ rungs 
takes the simple the form,

\begin{equation}
|N \rangle = \prod_{k=1}^N (k, \overline{k})  
\label{2}
\end{equation}

\noindent This state is actually 
the exact ground state of the rung Hamiltonian
defined by,

\begin{equation}
H_{\rm rung} = 
J'  \sum_{k=1}^N {\bf S}_k \cdot {\bf S}_{\overline{k}} 
\label{3}
\end{equation}

\noindent where the vertical coupling 
constant $J'$ is antiferromagnetic (i.e. $J' > 0$ ).

\vspace{30 pt}
\begin{figure}[h]
\begin{center}
\leavevmode
\epsfbox{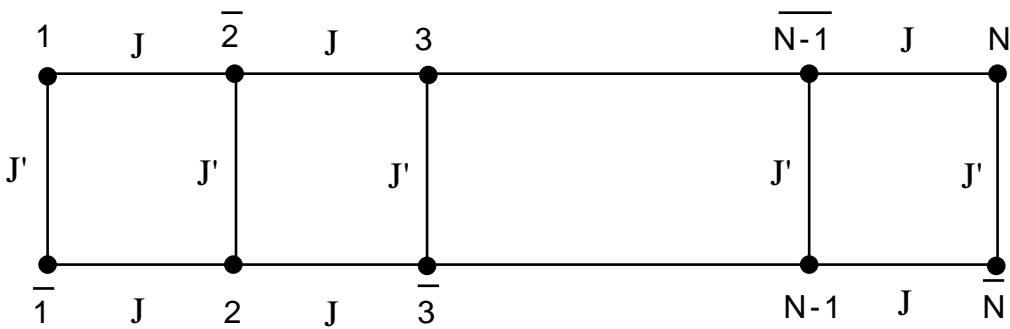}
\caption{A two-legged ladder showing the labels 
corresponding to sublattices A and B.}
\end{center}
\end{figure}

The correlation length $\xi$ of the state (\ref{2}) measured by 
the spin-spin correlator along the legs is exactly 0.

\noindent A trivial observation is that the states $|N+1 \rangle$ 
and $|N \rangle $
given by eq. (\ref{2}) are related by the equation,

\begin{equation}
|N+1 \rangle = |N \rangle \otimes |\phi_1 \rangle_{N+1} \label{4}
\end{equation}

\noindent where $ |\phi_1 \rangle_{N+1}$ denotes the singlet
state located at
the $(N+1)-$rung, namely,

\begin{equation}
 |\phi_1 \rangle_{N+1} = (N+1,\overline{N+1}) \label{5}
\end{equation}

\noindent Eq.(\ref{4}) 
is a $1^{st}$ order RR (i.e. $\nu = 1$), which upon $N-1$ iterations
reproduces the state (\ref{2}), 
given the initial condition $|1\rangle = |\phi_1\rangle$.


\subsection{Dimer State}

Let us assume that $h_0 = 1$ and $h_1$ are both
different from zero. The variational
parameter $h_1$ gives  
the amplitude of a horizontal bond in the RVB ansatz 
(\ref{1}), i.e.

\begin{equation}
 |N\rangle = \sum_{\overline{k}_{\alpha}} h_1^{n_1} 
(1, \overline{k}_1) (2, \overline{k}_2) \cdots (N, \overline{k}_N) \label{6}
\end{equation}

\noindent where $n_1$ is 
the number of horizontal bonds. The state (\ref{6}) is a 
linear superposition of dimer states of the form depicted
in figure 2, 
and for this reason is called a dimer-RVB
state. 
Observe that all the horizontal bonds come in pairs, 
for each horizontal bond
at one leg forces the presence of a 
companion in front of it at the opposite leg.

\vspace{30 pt}
\begin{figure}[h]
\begin{center}
\leavevmode
\epsfbox{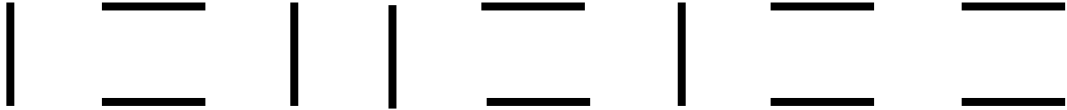}
\caption{A typical dimer state in a 2-legged ladder.}
\end{center}
\end{figure}

It is apparent that upon switching the Hamiltonian
which contains the couplings along the legs of the ladder, namely

\begin{equation}
H_{\rm leg} = J  \sum_{k=1}^{N-1}
( {\bf S}_k \cdot {\bf S}_{\overline{k+1}} + 
{\bf S}_{k+1} \cdot {\bf S}_{\overline{k}} ) \label{7}
\end{equation}

\noindent 
the dimer states 
(\ref{6}) will become more probable than the columnar state
(\ref{2}) which contains 
only vertical rungs, and for $J = J'$ one expects
the amplitude $h_1$  to be close to 1.
This expectation is based on the resonant mechanism that motivates the whole
RVB approach \cite{An}, \cite{Kivelson}, 
and which is the main cause of the 
substantial lowering of the energy for the RVB states (see figure 3).

The key observation for our purposes 
is that the dimer state (\ref{6}) of a ladder with
open boundary conditions   can be generated
from a $2^{nd}$ 
order RR given by:

\begin{equation}
|N+2\rangle = |N+1\rangle \otimes |\phi_1\rangle_{N+2} + 
u \; |N\rangle \otimes |\phi_2\rangle_{N+1,N+2} \label{8}
\end{equation}

\noindent where the state denoted by $ |\phi_2 \rangle_{N+1,N+2}$ 
is made up of a pair of horizontal  
bonds located between the rungs at $(N+1,N+2)$, i.e.,

\begin{equation}
|\phi_2 \rangle_{N+1,N+2} = 
(N+1,\overline{N+2})  (N+2,\overline{N+1}) \label{9}
\end{equation}

\noindent 
Comparing (\ref{6}) with (\ref{8}) we obtain the relation
between the parameters $h_1$ and  $u$,

\begin{equation}
u = h_1^2  \label{10}
\end{equation}

\vspace{30 pt}
\begin{figure}[h]
\begin{center}
\leavevmode
\epsfbox{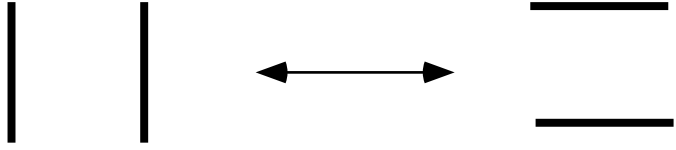}
\caption{The basic bond resonance mechanism
for an elementary plaquette.}
\end{center}
\end{figure}

\noindent
We refer to figure 4 
for a diagrammatic selfexplanatory representation of the
RR (\ref{8}).

One expects the dimer state (\ref{6}) 
to describe correctly ground states
for which the correlation length 
is at most one. 

\vspace{30 pt}
\begin{figure}[h]
\begin{center}
\leavevmode
\epsfbox{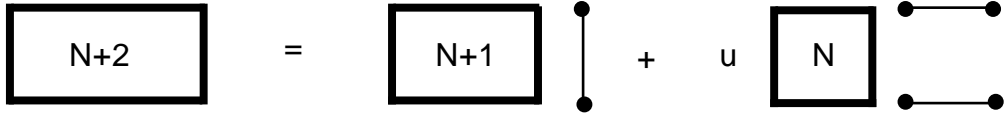}
\caption{Diagrammatic representation of the 
Recurrent Relation for $\nu=2$.}
\end{center}
\end{figure}

Fan and Ma have computed the  ground state energy and  
the number of dimer states contained
in the state $| N \rangle$ \cite{fan-ma}. 
Let us call that  number 
$F_N$ for reasons that will become clear below. $F_N$ is given by,

\begin{equation}
F_N = \frac{1}{ \sqrt{5}} ( {\alpha}_1^{N+1} - {\alpha}_2^{N+1})  
\label{11}
\end{equation}

\noindent where $\alpha_{1,2} = {1\over 2} (1 \pm \sqrt{5})$.
The key fact is to recognize eq.(\ref{11}) as  the Binet's 
formula  for the $N^{th}$ Fibonacci number \cite{Dic}. 
This last result inmediately follows 
from the RR satisfied by $F_N$, as a consequence of
eq.(\ref{8}), namely,

\begin{equation}
F_{N+2} =  F_{N+1} + F_N  \label{12}
\end{equation}

\noindent This equation together with the initial values,

\begin{equation}
F_0 =  F_1  = 1\label{13}
\end{equation}

\noindent reproduce  the well known  Fibonacci  sequence,

\begin{equation}
F_N = 1,1,2,3,5,8, \ldots \label{14}
\end{equation}

Another way to arrive to this result is 
by  counting the number of dimer
states in (\ref{6}). Calling   $M$ the number of pairs of horizontal 
bonds ( $M=0,1, \ldots , [N/2]$, where $[x]$ denotes the 
integer part of $x$), one gets,

\begin{equation}
F_N = \sum_{M=0}^{[N/2]} 
\left( \begin{array}{c} N-M \\ M \end{array} \right)
\label{15}
\end{equation}

\noindent
This formula is well known in the theory of partitions \cite{andrews}.

\subsection{Nondimer States}

These states are obtained whenever $h_l$ is non-zero for bonds connecting
non NN (nearest-neigbour) states. 
As an example, let us assume that the  
non vanishing   amplitudes are 
$h_0=1$, $h_1$ and $h_2$. In figure 5 we show 
``local" structures formed with these  bonds.
The  number  of configurations of this type
increases enormously with the number  of rungs.
However  not all of them are 
independent. The dimer states 
are all linearly independent.

\vspace{30 pt}
\begin{figure}[h]
\begin{center}
\leavevmode
\epsfbox{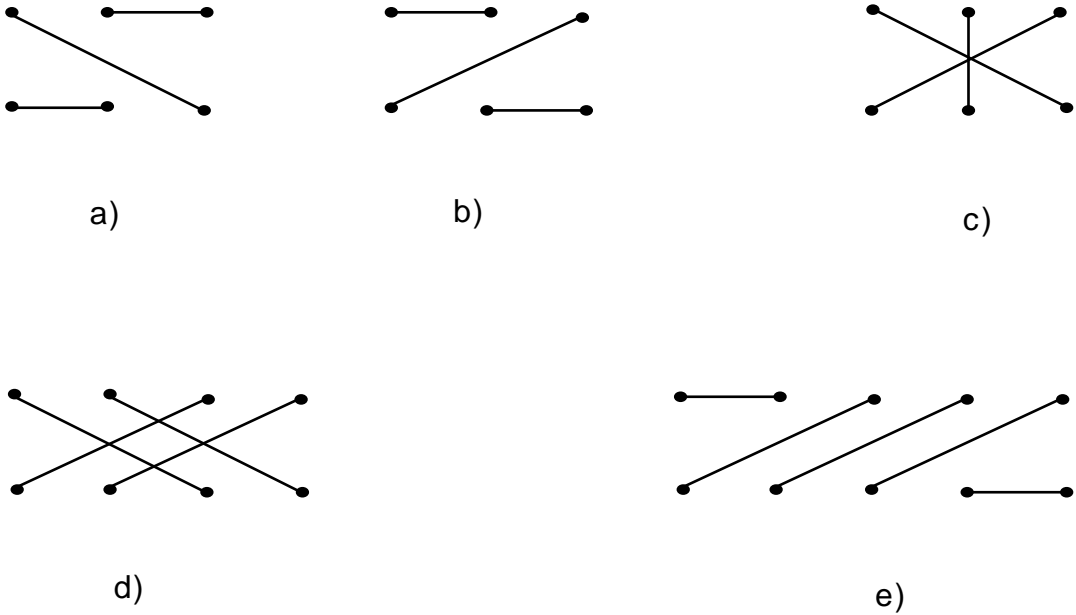}
\caption{A picture of several non-dimer states
showing their increasing variety.}
\end{center}
\end{figure}

\noindent For example, the state depicted 
in fig. 5c) can be written as a linear combination
of dimer states and the states in fig. 5a) and fig. 5b)
\cite{Oguchi} (see figure 6).

\vspace{30 pt}
\begin{figure}[h]
\begin{center}
\leavevmode
\epsfbox{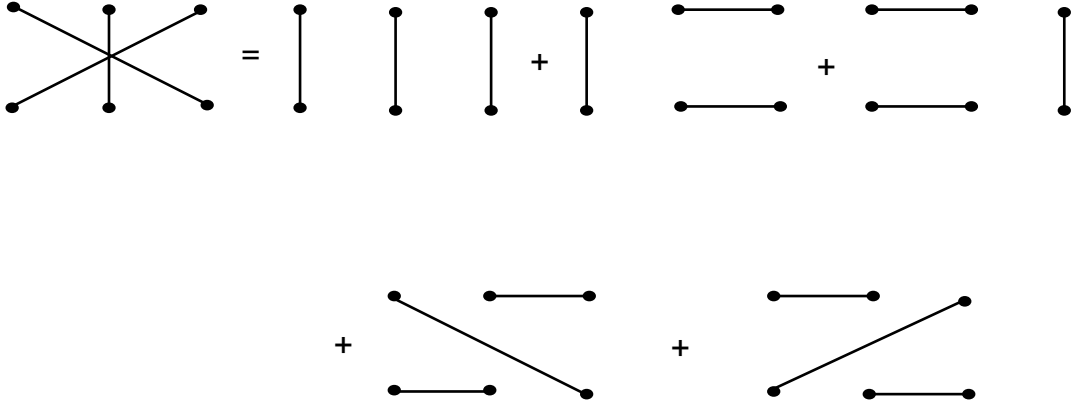}
\vspace{20 pt}
\caption{A non-dimer state as a linear combination 
of the 5 linearly independent states defining the
$\nu=3$ Recurrent Relation.}
\end{center}
\end{figure}

It is clear from figures 5 that in order to generate the nondimer RVB states
in full generality one has
to consider RR's with an arbitrary 
high order $\nu$. Thus, configurations 5a), 5b) and
5c) require a $3^{rd}$ order RR, 
configuration 5d) a $4^{th}$ order configuration, and
so on. For physical reasons one expects 
$h_2 <  h_0, h_1 $, so that configurations
5d) and 5e) are much less probable than configurations 5a),b),c).
Hence, if we restrict ourselves 
to these latter class of configurations we 
can again generate that class iteratively by means of the following RR,

\begin{equation}
|N+3\rangle = |N+2\rangle \otimes |\phi_1\rangle_{N+3} + 
u |N+1\rangle \otimes |\phi_2\rangle_{N+2,N+3} +
v |N \rangle \otimes |\phi_3\rangle_{N+1,N+2,N+3} 
\label{16}
\end{equation}

\vspace{30 pt}
\begin{figure}[h]
\begin{center}
\leavevmode
\epsfbox{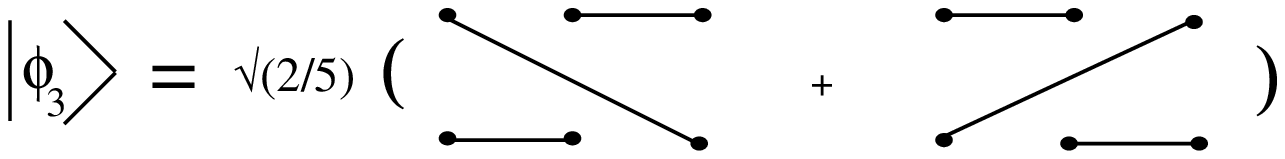}
\vspace{20 pt}
\caption{A pictorical representation of the
non-dimer state $|\phi_3 \rangle$ defining the
$\nu=3$ Recurrent Relation.}
\end{center}
\end{figure}

\noindent where $|\phi_3\rangle_{N+1,N+2,N+3} $ 
is pictured in figure 7
and its precise expression reads,

\begin{equation}
 |\phi_3\rangle_{1,2,3} = \sqrt{{2\over 5}} 
[ (1,\overline{2}) (2,\overline{3}) (3,\overline{1}) + 
(1,\overline{3}) (2,\overline{1}) (3,\overline{2}) ]
\label{17}
\end{equation}

\noindent The  prefactor $\sqrt{{2\over 5}} $ is for normalization.
The last term in (\ref{16}) generates states with two horizontal
bonds of type $h_1$ and one bond of type $h_2$,  
which follows the chess knight's move \cite{Iske}
( this kind of bonds
have been recently considered in the study of the Hubbard model
on $2 \times 2 \times 2$ and $4 \times 4$ lattices
in ref.\cite{fano} ). The relation between
$h_1, h_2$ and $v$ is given by,

\begin{equation}
\sqrt{ \frac{2}{5} } v = h_1^2 \; h_2
\label{18}
\end{equation}

We have not included 
in $|\phi_3\rangle$ the configuration 5c) since it
depends linearly on the remaining configurations contained in  (\ref{16}).
In a sense eq.(\ref{16}) generates recurrently the most important RVB
configurations with bonds of type $h_0$, $h_1$ and $h_2$.
The RR (\ref{16}) will  be used to generate all the states 
$|N \rangle$ with $N \geq 1$ provided we make the following
formal identifications,

\begin{equation}
|N \rangle = \left\{ \begin{array}{cc} 1, & N=0 \\ 0, & N < 0
\end{array} \right.
\label{formal}
\end{equation} 

These are also the initial conditions of the RR (\ref{8}).

The states generated recurrently 
from equations similar to (\ref{8}) and (\ref{16})
shall be called hereafter 
RVA states (standing for Recurrent Variational Approach), to 
distinguish them from the RVB states 
of the form (\ref{1}). We have proved above  
that the RVA states produced by
$1^{\rm st}$ and $2^{\rm nd}$ order RR's 
actually coincide with 
RVB states, but this is not true for higher order RVA states.
The main advantage in working with RVA states is that they
can be treated  analytically.

It is quite clear that we can 
perform some generalizations of the previous ideas.
The general form of a RR of order $\nu$ can be written as,

\begin{equation}
|N+\nu\rangle = \sum_{l=1}^{\nu} u_l |N+\nu-l\rangle \otimes 
|\phi_l\rangle_{N+\nu-l+1, \ldots ,N+\nu}
\label{19}
\end{equation}

\noindent where $|\phi_l\rangle$ 
are normalized states which must be chosen to be 
linearly independent from 
the states generated in the previous steps, while $u_l$
are variational parameters.

The RR satisfied by  the 
number, $F_N^{(\nu)}$, 
of linearly independent states generated by (\ref{19}), 
is given by:

\begin{equation}
F_{N+\nu}^{(\nu)} = \sum_{l=1}^{\nu} F_{N+\nu-l}^{(\nu)}
\label{20}
\end{equation}

\noindent subject to the initial conditions 
$F_0 = 1$ and  $F_N =0$ for $N <0$. 
Therefore the higher order RR's correspond to 
generalizations of the Fibonacci numbers. In 
section IV  we shall get  the  following 
Binet's formula for $F_N^{(\nu)}$, 

\begin{equation}
F_{N}^{(\nu)} = \sum_{i=1}^{\nu} \prod_{j\neq i} 
{\alpha_i^{N+\nu-1}\over \alpha_i - \alpha_j} \label{21}
\end{equation}

\noindent where $\alpha_i$ are the roots of the polynomial
$y^{\nu} - y^{\nu-1}- \ldots -y-1$.


\section{Recurrence Relations for Overlaps}

\indent In this and the following 
section we shall compute the value of the energy
$\langle N| H |N\rangle / \langle N|N \rangle$ of the  
RVA states $|N \rangle$.

Let us proceed progresively and first consider the $2^{nd}$ order RR given by 
eq.(\ref{8}). It is convenient to introduce the quantities $Z_N$ and $Y_N$ as:

\begin{equation}
\begin{array}{rl} 
Z_N = &  \langle N|N \rangle \\
Y_N = & _{N}\langle \phi_1| \otimes \langle N -1| N \rangle
\end{array}  \label{22} 
\end{equation}

\noindent Now one can easily derive the following RR's,

\begin{equation} 
 \begin{array}{rl}
Z_{N+2} = &  Z_{N+1} + u Y_{N+1} + u^2 Z_N \\
Y_{N+2} = &  Z_{N+1} + {u\over 2} Y_{N+1} \end{array}
\label{23} 
\end{equation}

\noindent where we have made use of the result,

\begin{equation}
_{N+1}\langle \phi_1 | \phi_2 \rangle_{N,N+1} = {1\over 2} |\phi_1 \rangle_N
 \label{24}
\end{equation}

The RR's (\ref{23}) together with the initial conditions,

\begin{equation}
\begin{array}{c} 
Z_{0} = Z_{1} = 1 \\
Y_{0} = 0, \  Y_{1} = 1 \end{array}
\label{25}
\end{equation}

\noindent determine 
$Z_N$ and $Y_N$ for arbitrary values of $N$.
This will be done in  the next section using generating-function
methods.

The Hamiltonian $H_N$ of a ladder with
$N$ rungs, is given by 

\begin{equation}
H_N = H_{\rm leg} + H_{\rm rung}
\label{26}
\end{equation}

\noindent where $H_{\rm leg}$ and $H_{\rm rung}$ 
are defined  in (\ref{3}) and (\ref{7}) respectively,
The RR method applied to Hamiltonian overlaps requires
the following definitions,

\begin{equation} 
\begin{array}{cl}
E_N = &  \langle N| H_N |N \rangle \\
D_N = & _{N}\langle \phi_1| \otimes \langle N -1| H_N |N \rangle
\end{array}
\label{27}
\end{equation}

\noindent A straightforward 
computation using eqs. (\ref{8}) and (\ref{24})
leads to the following RR's for $E_N$ and $D_N$,

\begin{equation} 
\begin{array}{cl}
E_{N+2} = & E_{N+1} + J' \epsilon_0 Z_{N+1} +
u ( D_{N+1} + (2J+J') \epsilon_0 Y_{N+1} ) + 
u^2 (E_N + 2J\epsilon_0 Z_N) \\
D_{N+2} = &  E_{N+1} + J' \epsilon_0 Z_{N+1} +
 {u\over 2} ( D_{N+1} + (2J+J') \epsilon_0 Y_{N+1} ) 
\end{array}
\label{28}
\end{equation}

\noindent where $\epsilon_0 = -3/4$ is the lowest eigenvalue of the operator 
${\bf S}_1 \cdot {\bf S}_2$. 

The initial conditions for $E_N$ and $D_N$ are,

\begin{equation} 
\begin{array}{c}
E_{0} = 0, \  E_{1} = J' \epsilon_0 \\
D_{0} = 0, \  D_{1} = J' \epsilon_0
\end{array}
\label{29} 
\end{equation}

Before showing the power of the RR method by computing the  values of 
several physical quantities, 
we set off for the generalization of equations (\ref{22})
and (\ref{27}) to arbitrary values of the order $\nu$ of the RR.
Let us define  the following quantities:

\begin{equation} 
\begin{array}{lc}
Z_{N,l} =  _{N,\ldots,N-l+1}\langle  
\phi_l| \otimes \langle  N - l |N \rangle & \\
 &  (l = 0,1,\ldots ,\nu-1) \\
E_{N,l} =   _{N,\ldots,N-l+1}\langle  
\phi_l| \otimes \langle  N - l | H_N |N \rangle & 
\end{array} 
\label{30} 
\end{equation}

\noindent where we identify $Z_{N,0} \equiv \langle N | N \rangle$ and 
$E_{N,0} \equiv \langle N| H_N | N \rangle$.

\noindent The RR's (\ref{19}) for the ground states $|N \rangle$, 
does not automatically imply that 
the overlaps $Z_{N,l}$  also satisfy RR's. However this is guaranteed
provided  the states 
$\{ |\phi_l \rangle  \}_{l=1}^{\nu}$ satisfy the following eqs.

\begin{equation}
_{N,\ldots,N-k+1}\langle \phi_k | \phi_l \rangle_{N-l+1,\ldots,N} = 
\left\{ \begin{array}{ll}
\Omega_{kl} \; |\phi_{l-k} \rangle_{N-l+1,\ldots,N-k} & l > k \\
\Omega_{ll} = 1 & l = k \\
\Omega_{kl} \;  _{N -l,\ldots,N-k+1}\langle \phi_{k-l} |   & l < k
\end{array} \right.
 \label{31}
\end{equation}

\noindent As a matter of fact, 
eq.(\ref{24}) gives an example of 
conditions (\ref{31}). The states 
$|\phi_1\rangle$, $|\phi_2\rangle$  and $|\phi_3\rangle$
also satisfy eqs. (\ref{31}),
with the overlapping matrix $\Omega$ given by,

\begin{equation}
\Omega^{(\nu=3)} = \left( \begin{array}{ccc}
1 & {1\over 2} & \sqrt{{2\over 5}} \\
{1\over 2} & 1 & \sqrt{{2\over 5}} \\
\sqrt{{2\over 5}}  & \sqrt{{2\over 5}}  & 1
\end{array} \right)
\label{32}
\end{equation}

\noindent Eqs.(\ref{32}) can be given  a geometrical meaning in terms 
of the disentangling of connected bonds ( see figure 8).

For $\nu=3$, beside the overlaps defined in eqs.(\ref{30}), 
we need in addition the following matrix element,

\begin{equation}
W_N = _{N}\langle \phi_1| \otimes _{N-1} \langle \phi_1| \otimes
\langle N-2| N \rangle
\label{33}
\end{equation}

The RR's for the overlaps $Z_{N,l} (l = 0,1,2)$ and 
$W_N$  can be derived from eq.(\ref{16}),

\begin{equation}
\begin{array}{cl}
Z_{N+3}= & Z_{N+2} + u^2 Z_{N+1} + v^2 Z_N + 2 u \Omega_{12}
Z_{N+2,1} + 2 v \Omega_{13} Z_{N+2,2} + 2 u v \Omega_{23} Z_{N+1,1} \\
Z_{N+3,1}= & Z_{N+2}+ u \Omega_{12} Z_{N+2,1} + 
v \Omega_{13} Z_{N+2,2}\\
Z_{N+3,2} = & u Z_{N+1} + \Omega_{12} Z_{N+2,1} + v \Omega_{23} 
Z_{N+1,1} \\ 
W_{N+3} =  & u \Omega_{12} Z_{N+1} +  Z_{N+2,1} + v\Omega_{12} \Omega_{13} 
Z_{N+1,1} \end{array}
\label{34}
\end{equation}

The initial data to solve these RR's are,

\begin{equation}
\begin{array}{llll}
Z_0 = 1 & Z_{0,1} = 0 & Z_{0,2} = 0 & W_0 = 0 \\
Z_1 = 1 & Z_{1,1} = 1 & Z_{1,2} = 0 & W_1 = 0 \\
Z_2 = 1+u+u^2 & Z_{2,1} = 1+ \frac{u}{2} & Z_{2,2} = \frac{1}{2}
+ u  & W_2 =  1 + \frac{u}{2} \end{array}
\label{35}
\end{equation}

\vspace{30 pt}
\begin{figure}[h]
\begin{center}
\leavevmode
\epsfbox{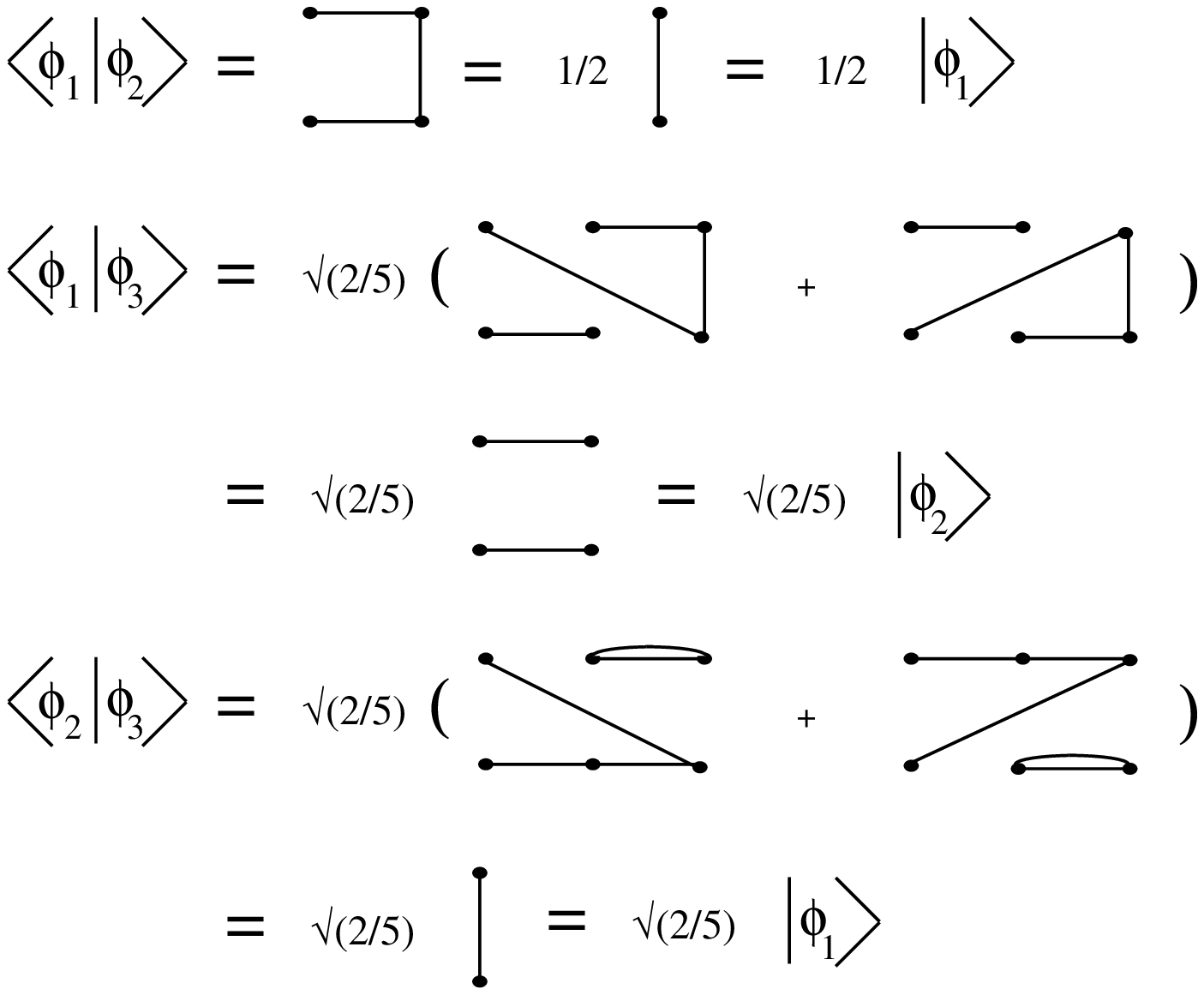}
\caption{Graphical representation of the 
overlapping calculus for the building 
block states.}
\end{center}
\end{figure}

In order to derive RR's for the overlaps of the Hamiltonian $H_N$
it is convenient to split $H_N$ as follows,

\begin{equation}
H_{N+3} = H_{N+l} + H_{N+l,N+3} 
\label{36}
\end{equation}

\noindent for different values of $l =0,1$ and 2 depending
on the particular matrix elements one needs to evaluate.
$H_{N+l, N+3}$ contains all the horizontal couplings between
the sites $N+l$ and $N+3$ and all the vertical couplings 
from  the rungs $N+l+1$ until $N+3$.

The analogue of eqs.(\ref{31}) are now,

\begin{equation}
\begin{array}{rl}
_{N+3, \dots, N+4-k} \langle \phi_k| H_{N+3-k,N+3} |\phi_k 
\rangle_{N+4-k, \dots, N+3} = & \epsilon_{kk} ,\,\,\, \;\; (k=1,2,3) \\
_{N+3}\langle \phi_1| H_{N+2,N+3} |\phi_2 
\rangle_{N+2, N+3} = & \epsilon_{12} | \phi_1 \rangle_{N+2} \\
_{N+3}\langle \phi_1 | H_{N+2,N+3} |\phi_3
\rangle_{N+1,N+2, N+3} = & \epsilon_{13} | \phi_2 \rangle_{N+1,N+2}
+ \epsilon'_{13} |\phi_1 \rangle_{N+1} |\phi_1 \rangle_{N+2} \\
_{N+3, N+2}\langle \phi_2| H_{N+1,N+3} |\phi_3 
\rangle_{N+1,N+2, N+3} = & \epsilon_{23} | \phi_1 \rangle_{N} 
\end{array}
\label{37}
\end{equation}

\noindent
together with their hermitian conjugated. The 
entries $\epsilon_{kl}$ can be collected into the  
symmetric $3 \times 3$ 
matrix,

\begin{equation}
 \epsilon_{kl}^{(\nu =3)}  = \epsilon_0 \left( \begin{array}{ccccc}
 J' & & J + {J'\over 2} & & \sqrt{{2\over 5}} ({2\over 3} J + J')  \\
 J + {J'\over 2} & & 2 J & & 3 \sqrt{{2\over 5}} \, J \\
  \sqrt{{2\over 5}} ({2\over 3} J + J')& & 
3 \sqrt{{2\over 5}} \, J & &{12\over 5}
(J + {1\over 4} J')
\label{38} \end{array}
\right) \end{equation}

\noindent while

\begin{equation}
\epsilon'_{13} =  \epsilon_0 \frac{2}{3} \sqrt{ \frac{2}{5} }J    
\label{39}
\end{equation}

The term proportional to $\epsilon'_{13}$ in (\ref{37})
is what forces us to introduce the overlap $W_N$ (\ref{33}).

The RR's for the energy overlaps are given by,

\begin{equation}
\begin{array}{cl}
E_{N+3} = & E_{N+2} + \epsilon_{11} Z_{N+2} + 2 u  ( \Omega_{12} 
E_{N+2,1} + \epsilon_{12} Z_{N+2,1} ) \\
&+ 2 v ( \Omega_{13} E_{N+2,2} + \epsilon_{13} Z_{N+2,2} + \epsilon'_{13}
W_{N+2} ) + u^2 ( E_{N+1} + \epsilon_{22} Z_{N+1} ) \\
& + v^2 ( E_N + \epsilon_{33} Z_N ) +
2 u v ( \Omega_{23} E_{N+1,1} + \epsilon_{23} Z_{N+1,1} ) \\
E_{N+3,1} = & E_{N+2} + \epsilon_{11} Z_{N+2} + 
u ( \Omega_{12} E_{N+2,1} + \epsilon_{12} Z_{N+2,1} ) \\
&+ v ( \Omega_{13} E_{N+2,2}  + \epsilon_{13} Z_{N+2,2} 
+ {\epsilon'}_{13} W_{N+2} \\
E_{N+3,2}= & \Omega_{12} E_{N+2,1} + \epsilon_{12} Z_{N+2,1} +
u ( E_{N+1} + \epsilon_{22} Z_{N+1} ) + v ( \Omega_{23}
E_{N+1,1} + \epsilon_{23} Z_{N+1,1} ) \end{array}
\label{40}
\end{equation}

The initial data for the energy overlaps are

\begin{equation}
\begin{array}{lll}
E_0 = 0 & E_{0,1} = 0 & E_{0,2} = 0  \\
E_1 = \epsilon_0 J'  & E_{1,1} = \epsilon_0 J' & E_{1,2} = 0  \\
E_2 = 2 \epsilon_0 [ J' + u(J+J') + u^2 J ] 
& E_{2,1} = \epsilon_0 [ 2 J' + u ( J'+J) ] 
 & E_{2,2} = \epsilon_0 ( J' + J + 2uJ) 
  \end{array}
\label{41}
\end{equation}


\section{Generating Function Methods for Solving the RR's}

The simplest way to find the general term of a series defined by a RR is by
introducing generating functions.
Fan and Ma in \cite{fan-ma} have also used generating functions 
in order to find the ground state energy of the dimer state. 
However in their approach they do not start from 
RR's that generate ground states, and so the appearance of 
generating functions seems rather obscure. 
Our method gives a simple and straighforward derivation
of their generating function methods.  
Let us illustrate the  technique with the derivation of the
Binet formula (\ref{11}).

\subsection{ Number of states}

For this purpose let us  define the 
following generating function,

\begin{equation}
F(x) = \sum_{N\geq 0} F_N x^N
 \label{42}
\end{equation}

A simple computation using (\ref{12}) and (\ref{13}) yields

\begin{equation}
F(x) = {1\over 1 - x - x^2}
 \label{43}
\end{equation}

\noindent To recover $F_N$ we can use countour integrals:

\begin{equation}
F_N = \oint_{C_0} {dx\over 2\pi i} x^{-N-1} F(x) \label{44}
\end{equation}

\noindent where $C_0$ is a countour that encloses the origin counterclockwise.

For reasons which will become clear 
later on, it is convenient to perform the change
of variables $x = 1/y$, in which case (\ref{44}) becomes:

\begin{equation}
F_N = \oint_{C_{\infty}} {dy\over 2\pi i} y^{N-1} \tilde{F}(y) = 
\sum Res (y^{N-1} \tilde{F}(y)) \label{45}
\end{equation}

\noindent where 

\begin{equation}
 \tilde{F}(y) \equiv F({1\over y}) \label{46}
\end{equation}

\noindent In (\ref{45}) $C_{\infty}$ is  
a contour around infinity which  picks up all the 
poles of the function 
$\tilde{F}(y)$. In the example (\ref{43}) we get,

\begin{equation}
 \tilde{F}(y) = {y^2 \over y^2 - y -1} \label{47}
\end{equation}

\noindent Noticing that $\alpha_{1,2}= 
{1\over 2} (1\pm \sqrt{5})$  are the two
roots of the equation $y^2 - y -1 = 0$, we get the Binet formula (\ref{11})

We can similarly derive 
eq.(\ref{21}) for the number $F_N^{(\nu)}$ of states
generated by the RR of order $\nu$. 
The generating functions $F^{(\nu)}(x)$ and
$\tilde{F}^{\nu} (y)$ are given by:

\begin{equation} 
\begin{array}{cl}
F^{(\nu)}(x) = &  1/( 1 - x - x^2 - \ldots - x^{\nu}) \\
\tilde{F}^{\nu} (y) = & y^{\nu}/( y^{\nu} - y^{\nu-1} - \ldots - y - 1)
\end{array}
\label{48}
\end{equation}

\subsection{ Norm and Energy of the $\nu=2$ states}

Let us now solve the RR's for $Z_N$ and $E_N$ in the $\nu = 2$ case.
To this end we introduce the following generating functions

\begin{equation}
\begin{array}{ll}
Z(x) = \sum_{N\geq 0} Z_N x^{N}, & Y(x) = \sum_{N\geq 0} Y_N x^{N} \\
E(x) = \sum_{N\geq 0} E_N x^{N}, & \ \ D(x) = \sum_{N\geq 0} D_N x^{N}
\end{array} 
\label{49}
\end{equation}

\noindent From equations (\ref{23}), 
(\ref{25}), (\ref{28}) and (\ref{29}) we find

\begin{equation}
\begin{array}{rl} 
(1 - x -x^2u^2) Z(x) = &  1 + xu Y(x) \\
(1 - {xu\over 2}) Y(x) = &  x Z(x) \\
(1 - x -x^2u^2) E(x) = &  \epsilon_0 (x J' + 2 x^2u^2 J) Z(x) + 
\epsilon_0 xu (2J + J') Y(x) + xu D(x) \\
(1 - {xu\over 2}) D(x) = &  x [E(x) + 
\epsilon_0 J' Z(x) + {u\over 2} (2J +J') \epsilon_0 Y(x)]
\end{array} 
\label{50}
\end{equation}

\noindent Then eliminating $Y(x)$ and $D(x)$ 
in terms of $E(x)$ and $Z(x)$ we find,

\begin{equation}
Z(x) = \frac{A(x)}{B(x)} ,\;\; E(x) = \frac{C(x)}{B(x)^2}
\label{51}
\end{equation}

\noindent
where  $A(x), B(x) $ and $C(x)$ are polinomials
of degrees $a=1, b=3, c= 4$ respectively,
given by

\begin{equation}
\begin{array}{rl}
A(x) = &  1 - \frac{1}{2} xu \\
B(x) =&  1 - (1 + {1\over 2} u) x - 
(u^2 + {1\over 2} u) x^2 +
{1\over 2} x^3 u^3 \;\;\; \;  \\
C(x) = &  \epsilon_0 \; x  \;  [J' + x u  (2J+J' + 2u J) - 
x^2 u^2  ( {1 \over 4} J'  + 2u J) + 
{1 \over 2} x^3 u^4 J]
\end{array}
\label{52}
\end{equation}

We shall show below that  eqs.(\ref{51}) also hold for 
$\nu=3$, where $A, B$ and $C$ are 
polinomials of degrees 3, 6 and 9 respectively.
For both, $\nu =2 $ and 3, 
we have the  relation,

\begin{equation}
a+b = c
\label{53}
\end{equation}

We shall again make the change of variables $x= 1/y$,
defining  the new polinomials,

\begin{equation}
P(y) = y^a A(1/y), \; Q(y) = y^b B(1/y), \;  R(y) = y^c C(1/y)
\label{54}
\end{equation}

In the large $N$ limit, $Z_N$ and $E_N$ are given by,

\begin{equation}
Z_N \sim \alpha^{N-1-a+b} \frac{ P(\alpha) }{ Q'(\alpha)} , \;\;\; 
E_N \sim N \alpha^{N-1+2b-c} \frac{R(\alpha) }{ \alpha 
Q'^2( \alpha)} 
\label{55}
\end{equation}

\noindent
where $Q'(y)= d Q(y)/dy$ and $\alpha$ denotes the biggest root
of the polynomial $Q(y)$.

The density energy per site of the 2-legged ladder is finally given by,

\begin{equation}
e_{\infty} = \lim_{N \rightarrow \infty}\frac{1}{2N} \frac{E_N}{Z_N}
= \frac{ R( \alpha) }{ 2 \alpha Q'(\alpha) P( \alpha) }
\label{56}
\end{equation}

\noindent
where we have employed eqs.(\ref{53},\ref{55}).

\subsection{ Norm and Energy of the  $\nu =3$ states}

A feature of the $\nu=2$ and 3 RR's for the generating functions 
$Z_l(x)$ and $E_l(x)$ , which we  believe is 
valid for higher order RR's,  is that they can be written in the
following compact form,

\begin{equation}
\begin{array}{ll}
\sum_{m=0}^{\nu-1} K_{l m}(x) Z_m(x) = e_l , & \\
& (l= 0, \dots, \nu-1) \\
\sum_{m=0}^{\nu-1} K_{l m}(x) E_m(x) = 
\sum_{m=0}^{\nu-1} L_{l m}(x) Z_m(x), &  \end{array}
\label{57}
\end{equation}

\noindent
where $(e_0, e_1, \dots, e_{\nu-1}) = (1,0, \dots,0)$. If $K_{l m}(x)$
is an invertible matrix then the solution of these eqs. is given by,

\begin{equation}
Z_l(x) = K^{-1}_{l 0}, \;\; E_l= K^{-1}_{l m} L_{ m n} K^{-1}_{n 0}
\label{58}
\end{equation}

Calling $M_{l n}$ the $(l,n)$ minor of the matrix $K_{ l n}$,
then eqs(\ref{58}) have the generic form 
postulated in (\ref{51}) with the identifications,

\begin{equation}
A(x) = M_{0 0}, \; \; B(x) = {\rm det} K(x), \;\; 
C(x) = M_{0 k} \; L_{k l} \; M_{l 0}
\label{59}
\end{equation}

The matrices $K(x)$ and $L(x)$  for $\nu=3$ can be derived
from eqs.(\ref{34}) and (\ref{40}) and they read,

\begin{eqnarray}
& K(x)  = & \label{60} \\
& \left(  \begin{array}{ccccc}
1 - x - {u^2}\,{x^2} - {v^2}\,{x^3}& &
- 2 u x \left( \Omega_{12}  + 2 v {x} \Omega_{23} \right) & &
-2\,v\,x \Omega_{13} \\
-x & & 1 - {{u\,x}\Omega_{12}} & &
- \Omega_{13}\,v\,x  \\
- u\,{x^2}  & &
-x \left( \Omega_{12} + \Omega_{23} \,v\,{x} \right) & & 1 
\end{array} \right) & \nonumber 
\end{eqnarray}

\begin{eqnarray}
& \frac{1}{x} \, L(x)  = & \label{61} \\
& \left( \begin{array}{ccccc}
\epsilon_{11}+\epsilon_{22} x u^2 +x^2 v( v \epsilon_{33}+
+ 2 u  \Omega_{12} \epsilon'_{13}) & &
2[ u \epsilon_{12} + u v \epsilon_{23} x +
v \epsilon'_{13}( x+\Omega_{12}\Omega_{13}v x^2) ] & &
2 v \epsilon_{13} \\
\epsilon_{11} + u v x^2 \Omega_{12} \epsilon'_{13} & &
u \epsilon_{12} + v \epsilon'_{13}(x + \Omega_{12} \Omega_{13} v x^2) & &
v \epsilon_{13} \\
u x \epsilon_{22} & & \epsilon_{12} + v x \epsilon_{23} & & 0 
\end{array} \right) & \nonumber 
\end{eqnarray}

\section{Minimization of the Ground State Energy: Results}

In the previous section we have derived
a formula for the ground state energy density, eq.(\ref{56}), in terms
of three polinomials evaluated at the biggest root $\alpha$
of  $Q(y)$. The minimization procedure consist now
in looking for the minimum   of $e_{\infty}$ by varying the parameter
$u$ for
$\nu=2$,
and  the parameters  $u$ and $v$ for $\nu=3$. 
In tables 1 and 2 we  show
the ground state energies per site for different values
of the coupling constant ratio $J/J'$, 
varying through strong, intermediate and weak coupling
regimes. The values $-e_{\infty}^{(2)}/J'$ and $-e_{\infty}^{(3)}/J'$
are those of the RVA states  for $\nu=2$ and 3.
The values  $-e_{\infty}^{MF}/J'$ are Mean Field values taken 
from \cite{GRS}, \cite{Jor}
while $-e_{\infty}^{Lan}/J'$ are Lanzcos values taken from \cite{Barnes}.

\begin{center} 
\begin{tabular}{|c|c|c|c|c|} 
\hline 
$J/J'$ & $u$ & $-e_{\infty}^{(2)}/J'$&   
$-e_{\infty}^{MF}/J'$ & $-e_{\infty}^{Lan}/J'$\\  \hline \hline

  $0$ & $0$ & $0.375$  & $0.375000 $  & $$ \\ \hline 

  $0.2$ & $0.128521$ & $0.383114$  &  $0.382548$  & $$ \\ \hline 

  $0.4$ & $0.323211$ & $0.40835$  &  $0.405430$  & $$ \\ \hline 

  $0.6$ & $0.578928$ & $0.44853$  &  $0.442424$  & $$ \\ \hline 

  $0.8$ & $0.87441$ & $0.499295$  &  $0.489552$  & $$ \\ \hline 

  $1$ & $1.18798$ & $0.556958$  &  $0.542848$  & $0.578$ \\ \hline 

  $1.25$ & $1.58519$ & $0.63518$  & $0.614473$  & $0.6687$ \\ \hline 

  $1.66$ & $2.21853$ & $0.772172$  &$0.738360$  & $0.8333$ \\ \hline 

  $2.5$ & $3.39153$ & $1.06915$  &$1.002856$  & $1.18$ \\ \hline 

  $5$ & $5.9777$ & $1.99285$  & $$  & $2.265$ \\ \hline \hline 
\end{tabular}
\end{center}
\begin{center}
Table 1
\end{center}

\begin{center}
\begin{tabular}{|c|c|c|c|c|c|} 
\hline \hline
  \mbox{J/J'} & $u$ & $v$ & $-e_{\infty}^{(3)}/J'$ & 
$-e_{\infty}^{MF}/J'$ & $-e_{\infty}^{Lan}/J'$\\  \hline \hline

  $0$ &  $0$ & $0$ & $0.375$ & $0.375000$  & $$ \\ \hline 

  $0.2$ &  $0.105059$ & $0.0200031$ & $0.383195$ & $0.382548$  & $$ \\ \hline 

  $0.4$ &  $0.230463$ & $0.0911438$ & $0.409442$ & $0.405430$  & $$ \\ \hline 

  $0.6$ &  $0.399714$ & $0.222909$ & $0.45252$ & $0.442424$  & $$ \\ \hline 

  $0.8$ &  $0.626206$ & $0.42656$ & $0.507909$ & $0.489552$  & $$ \\ \hline 

  $1$ &  $0.91059$ & $0.716812$ & $0.571314$ & $0.542848$  & $0.578$ \\ \hline 

  $1.25$ &  $1.34146$ & $1.22022$ & $0.657551$&$0.614473$ &$0.6687$ \\ \hline 

  $1.66$ &  $2.19835$ & $2.41718$&$0.808438$&$0.738360$ & $0.8333$ \\ \hline 

  $2.5$ &  $4.34045$ & $6.26652$&$1.13384$&$1.002856$ & $1.18$ \\ \hline 

  $5$ & $11.2015$ & $24.0891$ & $2.13608$ & $$  & $2.265$ \\ \hline \hline 
\end{tabular}
\end{center}
\begin{center}
Table 2
\end{center}

It is also possible to minimize the energy of ladders
with finite length. In table 3 we show the results obtained
with the $\nu=3$ ansatz at the isotropic value $J/J' =1$.
$N$ denotes the number of vertical rungs.

\begin{center} 
\begin{tabular}{|c|c|c|c|} 
\hline \hline
 $N$ & $u$ & $v$ & $-e_{N}^{(3)}/J'$  \\  \hline \hline

  $3$ &  $0.74225$ & $0.624464$ & $0.521564$ \\ \hline 

  $4$ &  $0.998703$ & $0.612503$ & $0.533225$ \\ \hline 

  $5$ &  $0.90358$ & $0.696407$ & $0.541372$  \\ \hline 

  $6$ &  $0.921508$ & $0.64184$ & $0.545944$  \\ \hline 

  $7$ &  $0.930216$ & $0.667329$ & $0.549671$   \\ \hline 

  $8$ &  $0.917952$ & $0.67691$ & $0.552377$  \\ \hline 

  $9$ &  $0.920783$ & $0.675904$ & $0.554453$ \\ \hline \hline 
\end{tabular}
\end{center}
\begin{center}
Table 3
\end{center}

From these results we can make the following comments

\begin{itemize}

\item For any value of $J/J'$ there is always a solution 
which minimizes the energy. In other words, the dimer and non dimer
RVA states are physical acceptable in the whole range of couplings.

\item In the strong coupling regime $J/J' < 1 $ the RVA states
give a slightly better ground state energy than the mean field result.
This later state produces rather unphysical results for $J/J' >1$,
which does not occur in our case.

\item The nondimer state improves considerably the g.s. energy
as compared with the dimer state, especially for $J/J' \geq 1$. 
As $J/J'$ increase the ratio $v/u$ also increases and it is greater
than one for $J/J' \sim 1.5$. Later on we shall discuss in more detail
the interpretation of the values of $u$ and $v$ in the isotropic case.

\item In the weak coupling limit $J/J' >>1$ $u$ approaches asymptotically
the value 15.94, while the energy density approaches - 0.379 $J$. The
later value is much greater than the exact result -0.4431 given by the
Bethe ansatz solution for the decoupled spin 1/2 chains. It is however
curious that in the limit $J/J' \rightarrow \infty$ the variational
parameter $ u$ does not go to infinity, yielding a dimer state whose
energy is - 0.375 $J$. This shows once more that the resonance mechanism
always lowers the energy.

\end{itemize}

\section{ Spin Correlation Length}

The technique we have used in the previous sections to compute the g.s.
energy of the RVA states can be easily extended to the evaluation
of spin-spin correlators.

\subsection{ Dimer state}

We want to  compute the correlator $\langle {\bf S}_1 \cdot {\bf S}_R \rangle$
between the spin operators at the positions 1 and $R$ 
on the ladder ( we assume for simplicity that both sites are on the
sublattice $A$). As usual
we shall define two auxiliary quantities

\begin{equation}
\begin{array}{cl}
g_N (R) = & \langle N | {\bf S}_1 \cdot {\bf S}_R|N  \rangle \\
f_N (R) = & \langle \phi_1 | \otimes \langle N-1|
{\bf S}_1 \cdot {\bf S}_R |N  \rangle \end{array}
\label{62}
\end{equation}

From the  RR (\ref{8}) one finds,

\begin{equation}
\begin{array}{cl}
g_{N+2} ( R) = & g_{N+1}(R) + u^2 g_{N}(R) + u f_{N+1}(R) \\
f_{N+2}(R) = & g_{N+1}(R) + \frac{u}{2} f_{N+1}(R) \end{array}
\label{63}
\end{equation}

The initial conditions are given by,

\begin{equation}
\begin{array}{rl}
g_N(R) = 2 f_N(R) = & \left\{ \begin{array}{ll} \frac{3}{2}
\left( \frac{u}{2} \right)^{R-1} & N= R \\
0 & N < R \end{array} \right. \end{array}
\label{64}
\end{equation}

These eqs can be derived using the Sutherland's rules 
\cite{sutherland} which imply that
there is only one loop covering giving a contribution to the 
correlation when the spin operators are located at the two boundaries
of the chain. Defining the generating functions $g(x), f(x)$ as
usual we convert the RR's (\ref{62}) into,

\begin{equation}
\begin{array}{rl}
(1 - x - x^2 u^2 ) g(x) = & g_R(R) x^R + x u f(x) \\
(1 - \frac{1}{2} x u) f(x) = & f_R (R) x^R + x g(x) \end{array}
\label{65}
\end{equation}

Eliminating $f(x)$ we get,

\begin{equation}
g(x) = g_R(R) x^R  / B(x) \Longrightarrow 
g_N(R) \sim \alpha^{N-R+2} g_R(R)/Q'(\alpha) 
\label{66}
\end{equation}

\noindent
where $\alpha$ is the biggest root of the polynomial $Q(y)$.
The correlator is finally given for $R >>1$ by,

\begin{equation}
\langle {\bf S}_1 \cdot {\bf S}_R \rangle = {g_N(R) \over Z_N} 
\sim \frac{ 3  }{ 2 (\alpha - u/2 )} 
\left( \frac{u}{ 2 \alpha} \right)^{R-1}
\label{67}
\end{equation}

\noindent
which gives the expected exponential decaying behaviour with the distance.
The correlation length is then  given by the expression,

\begin{equation}
{\rm e}^{ 1/\xi} = \frac{ 2 \alpha}{u}
\label{68}
\end{equation}

\noindent
From these result  we observe that ${\cal L} = 
{\rm exp}(1/\xi)$ 
satisfies  the cubic equation,

\begin{equation}
u^3 {\cal L}^3 - (2 + u) u^2 {\cal L}^2 - (2 + 4 u ) u^2 {\cal L}
+ 4 u^3 = 0
\label{69}
\end{equation}

Setting  $u=1$ in (\ref{69}) we obtain the 
solution of ${\cal  L}= 4.201 472 $
which yields a correlation length $\xi = 0.696 652 $. 
This result has been obtained before by White et al. in 
reference \cite{wns}, \cite{white}.

\vspace{30 pt}
\begin{figure}[h]
\begin{center}
\leavevmode
\epsfbox{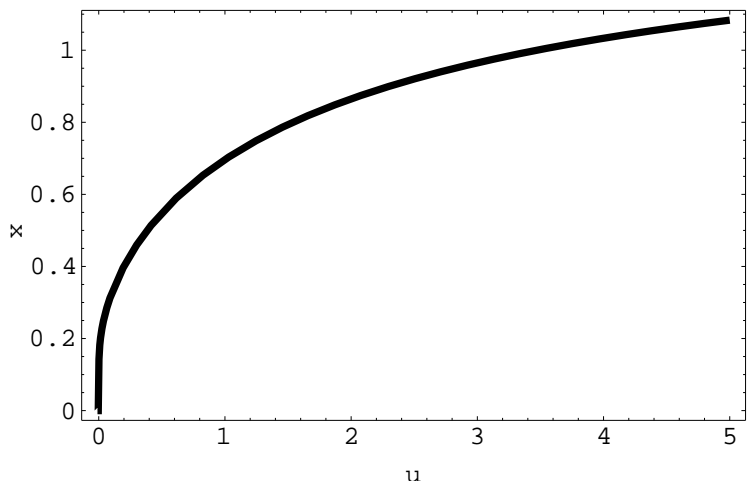}
\caption{The correlation length $\xi$ for the
dimer state $\nu=2$ as a function of the 
variational parameter $u$.}
\end{center}
\end{figure}

The correlation length as a function of the ratio 
$J/J'$ is found by replacing in (\ref{68}) the value of $u$ 
that minimizes the g.s. energy. We collect our results in Table 4.
For the isotropic case the g.s. corresponds to   $u= 1.18798$, which
gives  a the correlation length
$\xi =0.737 253 $, which is slightly bigger than 
the one given above for $u=1$.

\subsection{ $\nu=3$}

We shall outline the main steps of the
derivation of $\xi$, since they follow those of the
dimer case. In particular one gets the same asymptotic behaviour
$g_N(R) \sim \alpha^{N-R} g_R(R)$. The main difficulty of the nondimer
case is the computation of $g_R(R)$. Fortunately we can  apply 
some kind of ``nested" RR method 
to compute this quantity. In order to do so we need the
following definitions,

\begin{equation}
\begin{array}{rl}
g_R = & \langle R| {\bf S}_1 \cdot {\bf S}_R |R \rangle \\
g_{R,1} = & \langle \phi_1 | \otimes 
 \langle R-1 | {\bf S}_1 \cdot {\bf S}_R |R \rangle \\
g_{R,2} = & \langle \phi_2 | \otimes 
 \langle R-2 | {\bf S}_1 \cdot {\bf S}_R |R \rangle \\
\tilde{g}_{R,2} = & \langle \phi_1 | \otimes 
 \langle R-1 | {\bf S}_1 \cdot {\bf S}_{R-1} |R \rangle \end{array}
\label{70}
\end{equation}

The RR's satisfied by these quantities are given by,

\begin{equation}
\begin{array}{rl}
g_{R+3} = & 2 u \Sigma_{12}\; g_{R+2,1} + 2 v \Sigma_{13} 
\;(g_{R+2,2} + \tilde{g}_{R+2,2}) + 2 u v \Sigma_{23}\; g_{R+1,1} \\
g_{R+3,1} = & u \Sigma_{12}\; g_{R+2,1} + v \Sigma_{13} 
\;(g_{R+2,2} + \tilde{g}_{R+2,2})  \\
g_{R+3,2} = &  \Sigma_{12}\; g_{R+2,1} + v \Sigma_{23}\; g_{R+1,1} \\
\tilde{g}_{R+3,2} = &  \Omega_{12} \; g_{R+2,1} + v \Sigma_{23}\; g_{R+1,1} 
\end{array}
\label{71}
\end{equation}

\noindent
where we have use the matching eqs.

\begin{equation}
\begin{array}{rl}
_2\langle \phi_1 |{\bf S}_2 | \phi_2 \rangle_{1,2} = &
\Sigma_{12}\; {\bf S}_1 | \phi_1 \rangle_1 \\
_3\langle \phi_1 |{\bf S}_3 | \phi_3 \rangle_{1,2,3} = &
\Sigma_{13}\; ( {\bf S}_1 + {\bf S}_2 ) | \phi_2 \rangle_{1,2} \\
_{3,2}\langle \phi_2 |{\bf S}_3 | \phi_3 \rangle_{1,2,3} = &
\Sigma_{23}\; {\bf S}_1 | \phi_1 \rangle_1 \\
_{3,2}\langle \phi_2 |{\bf S}_2 | \phi_3 \rangle_{1,2,3} = &
\Sigma_{23} \; {\bf S}_1 | \phi_1 \rangle_1 \end{array}
\label{72}
\end{equation}

\noindent
with

\begin{equation}
\Sigma_{12} = 1/2 , \;\; \Sigma_{13} = \Sigma_{23} = \frac{1}{2}
\sqrt{ \frac{2}{5} } 
\label{73}
\end{equation}

The initial data to iterate (\ref{71}) are,

\begin{equation}
\begin{array}{llll}
g_1 = { 3/4}, & g_{1,1}= {3/4}, & g_{1,2} =0, & \tilde{g}_{1,2} = 0 \\
g_2 = { 3 u /4}, & g_{2,1}= {3 u/ 8}, & g_{2,2} =3/8, 
& \tilde{g}_{2,2} = \frac{3}{4} ( \frac{1}{2} +u)  \end{array}
\label{74}
\end{equation}

Using generating functions we get that the asymptotic behaviour of
$g_R$ is given by $\beta^R$ where $\beta$ is the highest root
of the following cubic polynomial,

\begin{equation}
\tilde{Q}(y) = y^3 - \Sigma_{12} u y^2 - 2 v \Sigma_{12}\; \Sigma_{13} \; y
- 2 v^2 \Sigma_{13} \; \Sigma_{23} \; 
\label{75}
\end{equation}

Hence the correlation length $\xi$ is given by 
the formula,

\begin{equation}
{\rm e}^{ 1/\xi} = \frac{ \alpha}{\beta}
\label{76}
\end{equation}

\noindent
which contains eq.(\ref{68}) as a particular case. Indeed
setting  $v=0$ in (\ref{75})  we get $\beta = u/2$.
 
In table 4 we show the values of $\xi$ computed from eq.(\ref{76}),
for those values of $u$ and $v$ that minimizes the g.s. energy.

\begin{center}
\begin{tabular}{|c|c|c|} 
\hline \hline
  $J/J'$ & $\xi^{(2)}$ & $\xi^{(3)}$ \\  \hline \hline

  $0$ & $0$  & $0$  \\ \hline 

  $0.2$  & $0.348297$  & $0.437166$  \\ \hline 

  $0.4$ & $0.471180$  & $0.608323$   \\ \hline 

  $0.6$ & $0.577543$  & $0.751286$   \\ \hline 

  $0.8$ & $0.665915$  & $0.866958$  \\ \hline 

  $1$ & $0.737253$  & $0.959249$   \\ \hline 

  $1.25$ & $0.807398$  & $1.04877$   \\ \hline 

  $1.66$ & $0.890758$  & $1.15205$  \\ \hline 

  $2.5$ & $0.994468$  & $1.26951$  \\ \hline 

  $5$ & $1.121437$  & $1.38532$   \\ \hline \hline 
\end{tabular}
\end{center}
\begin{center}
{Table 4}
\end{center}

Some comments are in order

\begin{itemize}

\item There is an improvement of the correlation length
specially when $J/J'$  increases. However the value of $\xi$ at the
isotropic point is still far from the 
numerical result which is 3.2 \cite{wns}, \cite{Greven}, \cite{Beat}. 
This means  that one should consider a RVA ansatz with longer bonds.

\item If for the dimer state we let $u$ to go to infinity,  then 
$\alpha \sim u$ and we get an upper limit for $\xi$

\begin{equation}
\xi^{(2)} \leq 1/ {\rm ln}2 = 1.442 695
\label{77}
\end{equation}

\item Setting $u=0$ and taking $v$ large one gets that $g_R \sim v^{2 R/3}$.
This behaviour can be understood diagrammatically by constructing the
bond configurations that contribute to the correlation 
$\langle {\bf S}_1 \cdot {\bf S}_R \rangle $, which are given by 
a succesion of states $\phi_1 \phi_3$, and such that the loop covering
generated by their overlap looks like a braid connecting the two
extremes of the ladder. The upper bond of $\xi$ is given in this case
by,

\begin{equation}
\xi^{(3)} \leq 3/ {\rm ln}5 = 1.864 008 , \,\,( u=0)
\label{78}
\end{equation}

\end{itemize}

\section{Discussion and Prospects}

The results of these paper give a confirmation of the short range
RVB picture of the 2-legged ladder proposed in \cite{wns}, especially
in the strong and intermediate coupling regimes.
Indeed, the state $\phi_3$  
can be considered
as a pair of two separated topological deffects 
connected by a long bond of type $h_2$ and two 
dimer bonds in a staggered
``high energy" configuration. 
The nondimer RVA  state generated by this type of local
configurations, 
improves the g.s. energy and correlation length of the dimer
state, but for $J \sim J'$ one needs to consider longer 
bonds in order to approach  the correlation length
computed using  QMC or DMRG methods.

It is interesting to evaluate the RVB parameters
$h_1, h_2$ which correspond to the RVA parameters $u$ and $v$.
According to eqs.(\ref{10}, \ref{18}) they are given, for the case
$J/J' =1$, by

\begin{equation}
h_1 = 0.954 248, \;\; h_2 = 0.497 865
\label{79}
\end{equation}

\vspace{30 pt}
\begin{figure}[h]
\begin{center}
\leavevmode
\epsfbox{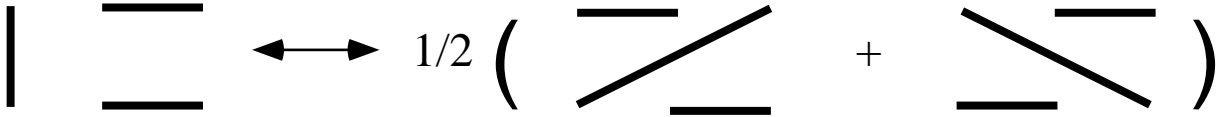}
\vspace{20 pt}
\caption{Mixed resonating mechanism involving the
building block states of the $\nu=3$ Recurrent Relation.}
\end{center}
\end{figure}

The value of $h_1 \sim 1 $ is due to the resonance mechanism
depicted in figure 3. On the other hand the remarkable 
proximity of
$h_2$ to $1/2$ can be interpreted in terms of the resonance between
horizontal and vertical  bonds having NN sites ( see figure 10).
This interpretation leads us to conjecture that for longer bonds
the RVB amplitudes should behave approximately  
for the isotropic ladder as,

\begin{equation}
{\rm For} \;\; J=J' ,\;\;{\rm and} \;\; n \geq 1, \;\;
 \; h_n \approx 1/ 2^{n-1}
\label{80}
\end{equation}

This guess of course agrees with the expected exponential
decaying behaviour of bond amplitudes of short range states \cite{LDA}.
It would be interesting to confirm (\ref{80}) either by higher
order RVA ansatzs or by Monte Carlo methods as those of ref \cite{LDA}.

The RR method can also be used to compute the value of the 
spin gap and the string order parameter which characterizes
the hidden topological LRO of the ladder. These results will be presented
elsewhere.
We have focused  in this paper on the two-legged spin ladder but
the techniques developped so far can in principle be 
applied  to  4, 6, \dots legged ladders, and more generally 
to systems with short range correlations as the spin 1 
Heisenberg chains, etc.
 The idea behind the RR method, has some similarities
with the  RG method of Wilson or the DMRG of White \cite{DMRG} in the sense
of constructing the ground state of a system in succesive steps
by  adding new sites (see also the variational formulation
of the DMRG  \cite{OR}). 
The role of $\nu$ is similar
to the role of the number of states $m$ kept in the DMRG method. 
For moderate values of the order $\nu$ of the RR we can perform
analytic computations of the g.s. energy and correlations. To improve the
accuracy of the RR method would require to implement it numerically. 
In summary the RR method gives us a   way to study low dimensional
quantum systems which is worth to pursue.

{\bf Acknowledgements} We would like to thank for conversations
to S. White, R. Noack, G. Fano and J. Dukelsky.
Work supported by
Spanish Grant AEN 96-1655 (G.S.) and  by
CICYT under  contract AEN93-0776
(M.A.M.-D.) .


\end{document}